\documentclass[journal]{IEEEtran}

\usepackage[noadjust]{cite}
\usepackage{soul}

\usepackage{hyperref}
\newcommand{\figref}[2][{}]{\hyperref[#2]{\figurename~\ref{#2}#1}} 

\newcommand{\andes}{A{\small NDES} }

\ifCLASSINFOpdf
   \usepackage[pdftex]{graphicx}
   \graphicspath{{./pdf/}{./jpeg/}}
   \DeclareGraphicsExtensions{.pdf,.jpeg,.png}
\else
   \usepackage[dvips]{graphicx}
   \graphicspath{{../eps/}}
   \DeclareGraphicsExtensions{.eps}
\fi

\usepackage{balance}

\usepackage{amsmath}
\usepackage{cleveref}
\usepackage{booktabs}
\usepackage{longtable}

\usepackage{array}

\ifCLASSOPTIONcompsoc
 \usepackage[caption=false,font=normalsize,labelfont=sf,textfont=sf]{subfig}
\else
 \usepackage[caption=false,font=footnotesize]{subfig}
\fi

\usepackage{stfloats}

\ifCLASSOPTIONcaptionsoff
\usepackage[nomarkers]{endfloat}
\let\MYoriglatexcaption\caption
\renewcommand{\caption}[2][\relax]{\MYoriglatexcaption[#2]{#2}}
\fi

\usepackage{listings}
\usepackage{courier}

\usepackage{url}

\begin{document}
\title{On the Modeling and Simulation of Anti-Windup Proportional-Integral Controller}

\author{Hantao~Cui,~\IEEEmembership{Member,~IEEE,}
  Yichen~Zhang,~\IEEEmembership{Member,~IEEE,} Federico
  Milano,~\IEEEmembership{Fellow,~IEEE,}
  Fangxing~(Fran)~Li,~\IEEEmembership{Fellow,~IEEE}%
  \thanks{This work was supported in part by the Engineering Research Center Program of the National Science Foundation and the Department of Energy under NSF Award Number EEC-1041877 and the CURENT Industry Partnership Program.}
  \thanks{H. Cui, and F. Li are with the Department of Electrical
    Engineering and Computer Science, The University of Tennessee,
    Knoxville, TN, 37996, USA. Email:
    fli6@utk.edu.}%
  \thanks{Y. Zhang is with Argonne National Laboratory, Lemont, IL,
    60439 USA.}%
  \thanks{F. Milano is with University College Dublin, Belfield, Ireland.}
}%

\markboth{Preprint to be submitted to IEEE Transactions on Circuits and 
Systems II: Express Briefs}%
{Cui \MakeLowercase{\textit{et al.}}: On the Modeling and Simulation}

\maketitle

\begin{abstract}
This paper investigates the chattering and deadlock behaviors of
the proportional-integral (PI) controller with an anti-windup (AW)
limiter recommended by the IEEE Standard 421.5-2016.
Depending on the simulation method, the controller may enter a 
chattering or deadlock state in some combinations of parameters and inputs.
Chattering and deadlock are analyzed in the context of three numerical
integration approaches: explicit partitioned method (EPM),
execution-list based method (ELM), and 
implicit trapezoidal method (ITM). This paper derives the
chattering stop condition for EPM and ELP, and
analyzes the impacts of step size and convergence tolerance for
simultaneous method. The deduced chattering stop conditions and 
deadlock behavior is verified with numerical simulations.
\end{abstract}


\IEEEpeerreviewmaketitle

\section{Problem Statement}

\begin{figure}[!t]
  \centering
  \includegraphics[width=0.65\columnwidth]{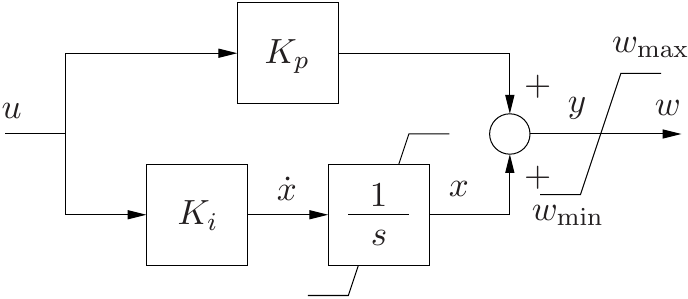}
  \caption{IEEE Standard 421.5-2016 PI model with conditional AW integrator.}
  \label{fig:pi-diagram}
  \vspace{-2mm}
\end{figure}

\IEEEPARstart{D}{iscontinuities} in power system simulations are
intricate problems that require careful handling. Anti-windup (AW) limiter
is one type of discontinuous component for modeling the saturation of devices.
The IEEE Standard 421.5-2016  \cite{lee1992ieee} recommends a 
proportional-integral (PI) controller block \cite{aastrom2006advanced,glattfelder2012control}
with an AW limiter \cite{kothare1994unified,tarbouriech2009anti,galeani2009tutorial}
as shown in \figref{fig:pi-diagram}, and
the recommended implementation is as follows:
\begin{equation}
  \begin{array}{lll}
    \text{if } y \ge w_{\max} &\Rightarrow \; \; \dot{x} = 0, \; \; w=w_{\max} \, , \\
    \text{if } y \le w_{\min} &\Rightarrow \; \; \dot{x} = 0, \; \; w=w_{\min} \, ,\\
    \text{otherwise} &\Rightarrow \; \; \dot{x} = K_i u, \; \; w=y, \; \; y=K_p u + x \, .
  \end{array}{}
  \label{eq:rec-implementation}
\end{equation}{}

The uniqueness of this model is that the AW limiter on the integrator
is conditional, depending on the hard limiter status. 
The differential-algebraic equation (DAE) formulation
introduces one differential variable $x$ and two algebraic variables,
$y$ and $w$, as follows:
\begin{equation}
  \begin{array}{lll}
    \dot{x} &= z_i K_i u \, ,\\
    0 &= (K_p u + x) - y \, ,\\
    0 &= (z_i y + z_l w_{\min} + z_u w_{\max}) - w \, .
  \end{array}{}
  \label{eq:dae-formulation}
\end{equation}

It is common to use boolean variables to implement the AW effect and
use a piecewise equation to set the hard limits \cite{murad2018impact,visioli2003modified}.
Variables $z_i$,
$z_u$ and $z_l$ represent within, hitting the upper limit, and hitting
the lower limit, respectively.

Chattering means excessive limiter switching in a finish-able
simulation, while deadlock means indefinite switching within a time
step that halts a simulation.  Depending on the controller parameters
and inputs, this implementation may cause chattering or deadlock in
simulations \cite{Murad2019}.  Qualitatively, consider the
proportional and the integrator outputs are in different directions
when the integrator is to about to be unlocked.  If the integrator
output exceeds the proportional one in magnitude, the hard limiter
will become binding. The integrator could be disabled and enabled back
and forth, causing issues until the proportional changes become
dominating.

\begin{table}
\caption{Equation sets depending on the entering limiter status}
\label{tab:conditional-equation-sets}
\centering
\begin{tabular}{ll}
\hline
$\mathbf{z_i = 0}$                             & $\mathbf{z_i = 1}$             \\
\hline
$\dot{x} = 0$                         & $\dot{x} = z_i K_i$   \\
$0 = (K_p u + x) - y$                 & $0 = (K_p u + x) - y$ \\
$0 = (z_l w_{\min} + z_u w_{\max}) - w$ & $0 = y - w$           \\
\hline
\end{tabular}
\end{table}

In fact, \eqref{eq:dae-formulation} can be split into two sets of
equations, based on the limiter state, given in
\Cref{tab:conditional-equation-sets}.  Chattering or deadlock
happens if the solutions do not satisfy the entering limiter status.
In such s scenario, the limiter status will toggle, and the other
set of equations will be solved.  Since the switching of equations
are discontinuous, when the limiter status keeps toggling, the
numerical solutions will jump back and forth.

Chattering or deadlock is determined by the applied numerical
integration method.  Two numerical integration methods are widely used
for power system simulation, namely, the explicit partitioned method
(EPM) and the implicit trapezoidal method (ITM).  Additionally, the
execution list-based method (ELM) adopted in Simulink™ is also
commonly used for block-based modeling and simulation.

This paper analyzes the chattering and deadlock behaviors under the
three above-mentioned integration methods.  \Cref{sec:workflow}
discusses workflows of integration methods and impacts on the PI
controller simulation.  \Cref{sec:partitioned} discusses the
chattering behavior using the EPM and ELM.  \Cref{sec:simultaneous}
discusses the deadlock using ITM and the impacts of step size and
convergence tolerance.  \Cref{sec:Conclusions} concludes the finding.

\section{Numerical Integration Workflows}
\label{sec:workflow}

As seen in \figref{fig:pi-diagram}, the integrator state variable is
dependent on the output $y$, an algebraic variable. The equation
evaluation workflow of integration methods will directly affect the
chattering and deadlock behaviors of the controller.
Three most commonly used methods for power system transient dynamic
simulation are discussed in the following.
\begin{enumerate}
\item \textit{Explicit Partitioned Method (EPM)}: algebraic equations
  that approximate fast electromagnetic transients are first solved,
  followed by the integration of differential equations using an
  explicit formula \cite{chow2019power}:
  \begin{equation}
    \label{eq:general-partitioned}
    \begin{array}{ll}
      0 = \mathbf{g}(\mathbf{x}_t, \mathbf{y}_{t+h}, \mathbf{u}_{t+h}) &
      \Rightarrow \text{solve for } \mathbf{y}_{t+h} \, , \\
      \dot{\mathbf{x}}_{t+h} = \mathbf{f}(\mathbf{x}_t, \mathbf{y}_{t+h}, \mathbf{u}_{t+h}) &
      \Rightarrow \text{solve for } \mathbf{x}_{t+h} \, ,
    \end{array}
  \end{equation}
  where $\mathbf{x}$, $\mathbf{y}$, and $\mathbf{u}$ are the states,
  algebraic variables, and discrete states, respectively.  To simulate
  from $t$ to $t+h$, the two equation sets in
  \eqref{eq:general-partitioned} are solved sequentially.
  Although iterations can be applied until $y_{t+h}$ and $x_{t+h}$ stop changing, 
  iterative implicit methods offer better numerical stability when iterative approaches are needed.
  The non-iterative EPM is the most commonly used method in commercial simulation tools. 
\item \textit{Implicit Methods}: the differential equations are solved
  along with algebraic equations:
  \begin{equation}
    \label{eq:general-simultaneous}
    \begin{aligned}
      0 &= \mathbf{g}(\mathbf{x}_{t+h}, \mathbf{y}_{t+h}, \mathbf{u}_{t+h}) \, , \\
      0 &= \dot{\mathbf{x}}_{t+h} - \mathbf{f}(\mathbf{x}_{t+h}, \mathbf{y}_{t+h}, \mathbf{u}_{t+h}) \, .
    \end{aligned}
  \end{equation}
  There are a variety of implicit methods utilized in power systems,
  being the implicit trapezoidal method (ITM) the most popular one
  \cite{milano2010power}.  The ITM applied to
  \eqref{eq:general-simultaneous} leads to:
  \begin{equation}
    \label{eq:itm}
    \begin{aligned}
      0 &= \mathbf{g}(\mathbf{x}_{t+h}, \mathbf{y}_{t+h}, \mathbf{u}_{t+h}) \, , \\
      0 &= \mathbf{x}_{t+h} - \mathbf{x}_t - 0.5 h (\mathbf{f}(\mathbf{x}_{t+h}, \mathbf{y}_{t+h}, \mathbf{u}_{t+h}) - \mathbf{f}_t) \, ,
    \end{aligned}
  \end{equation}
  where $\mathbf{x}_t$ and
  $\mathbf{f}_t = \mathbf{f}(\mathbf{x}_{t}, \mathbf{y}_{t},
  \mathbf{u}_{t})$ are the known vectors of state variables and
  differential equations, respectively, evaluated at step $t$.
  Due to robustness and generality, ITM is also widely adopted in
  both open-source \cite{milano2005open,cui2020hybrid} and commercial tools.
\item \textit{Execution List-based Method (ELM)}: algebraic and
  differential equations are solved sequentially for each
  block. Blocks are solved sequentially as defined in the execution
  list \cite{mathworks2020}. Equations are not grouped like in a power system simulation
  tool.  Rather, the evaluation sequence is based on the data flow
  specific to the model.
  For example, to integrate from $t$ to $t+h$ for the PI controller under discussion,
  ELM evaluates $\dot{x}_{t+h}$, $x_{t+h}$ (based on $z_{i,t}$),
  $y_{t+h}$ (with the corresponding $z_{i, t+h}$),
  and $w_{t+h}$ in sequence, based on \eqref{eq:dae-formulation}.
  ELM has the advantage of fully representing the control logic 
  and signal flow in digital controllers.
\end{enumerate}

It is worth mentioning that ELM and the non-iterative form of
EPM introduce a ``delay'' equal to the time step $h$ between
$\mathbf{x}$ and $\mathbf{y}$.  
Implicit methods are iterative for
nonlinear systems and can guarantee solutions to satisfy all equations
for each step, but they may show convergence issues for discontinuous
right-hand side equations, i.e.~when the equations model anti-windup
limiters.  Due to the workflow and stop criteria, chattering can
only happen when solved with non-iterative methods, while deadlock can
only occur with iterative ones.

Following sections study the conditions in which chattering or
deadlock can be stopped or avoided.  All following analyses start from
a generic time $t-h$ and consider a decreasing input ($u <0$ in the
proximity of $t-h$).  We also assume the output $y_{t-h}$ is at the
upper limit that initially locks the integrator.

\section{PI Controller Chattering}
\label{sec:partitioned}

\subsection{Chattering with Explicit Partitioned Method}

The most relevant characteristic of EPM to the PI controller is that
the change in output $y$ can unlock the integrator for the current
time step.  However, the change in state $x$ is only reflected at the
next time step.  Based on the assumptions in \Cref{sec:workflow},
initial conditions in EPM are given by:
\begin{equation}
  \label{eq:partitioned-method-t-h}
  \begin{array}{ll}
    y_{t-h} &= K_p u_{t-h} + x_{t-2h} = w_{\max} \, ,\\
    \Rightarrow &z_{i, t-h} = 0 \ \& \ x_{t-h} = x_{t-2h} \, .
  \end{array}
\end{equation}
At time $t$, as $u$ decreases, $y_t$ drops below the upper limit,
unlocking the integrator for time $t$, as given by:
\begin{equation}
  \label{eq:partitioned-method-t}
  \begin{array}{llll}
    y_{t} &= K_p u_{t} + x_{t-h} = y_{t-h} + K_p \Delta u_{t}  < w_{\max} \, ,\\
    \Rightarrow &z_{i, t} = 1 \ \&\ x_{t} = x_{t-1} + \Delta x_t \, ,
  \end{array}
\end{equation}
where the unlocked limiter will be reflected in $x_t$, as well as in
$y_{t+h}$ due to the ``delay'' in EPM.

For the subsequent time steps, if the output $y_{t+kh}$
($k=1, ..., N$) do not return to $w_{\max}$, the controller is
considered to have stopped chattering. For the immediate next step
$t+h$, the condition is given by \eqref{eq:partitioned-method-t+h}.
\begin{equation}
  \label{eq:partitioned-method-t+h}
  \begin{array}{lll}
    y_{t+h} &= y_{t-h} + K_p (\Delta u_t + \Delta u_{t+h}) + \Delta x_t < w_{\max} \, , \\
    \Rightarrow & K_p (\Delta u_t + \Delta u_{t+h}) + \Delta x_t < 0 \, .
  \end{array}
\end{equation}

Since $\Delta x_t$ is integrated from $\dot{x}_t$, which only depends
on $u_t$, $\Delta x_t$ will evaluate to $h K_i u_t$ regardless of the
integration formula.  The generalized condition for the subsequent
step $t+kh$, where $k$ is the number of steps ahead, is given by
\begin{equation}
  \label{eq:partitioned-method-chatter-free-t+ih}
  K_p \sum _{i=0}^k \Delta u_{t+ih} +h K_i (k u_t + \sum_{i=0}^{k-1} k\Delta u_{t+ih}) < 0 \, .
\end{equation}
Equation \eqref{eq:partitioned-method-chatter-free-t+ih} must hold for
a sufficient number of steps to avoid chattering.  Note that it
depends on the initial condition and trajectory of $u_t$ and is
case-specific.
When such condition is satisfied, the integrator will not become locked 
again after being unlocked.

\subsection{Chattering with Execution List-based Method}

ELM is different from EPM in the equation evaluation sequence. Since
the integrator outputs to the summation block, state $x_t$ is
integrated based on $z_{i, t-h}$, which is from the previous time step,
before calculating $y_t$.  The initial conditions for ELM at
$t-h$ are given by
\begin{equation}
  \label{eq:ELM-t-h}
  y_{t-h} = K_p u_{t-h} + x_{t-h} = w_{\max}, \text{and } z_{i, t-h} = 0 \, .
\end{equation}
At time $t$, the output $y_t$ is given by \eqref{eq:ELM-t}
\begin{equation}
  \label{eq:ELM-t}
  \begin{array}{cc}
    x_t = x_{t-h} \\
    y_t = y_{t-h} + K_p \Delta u_{t} < w_{\max}, \text{and } z_{i, t} = 1 \, .
  \end{array}
\end{equation}

Similarly, $y_{t+h}$ needs to remain smaller than $w_{\max}$.
\begin{equation}
  \label{eq:ELM-t+h}
  \begin{array}{lll}
    x_{t+h} &= x_t + \Delta x_{t+h} \, , \\
    y_{t+h} &= y_{t-h} + K_p (\Delta u_{t} + \Delta u_{t+h}) + \Delta x_{t+h} < w_{\max} \, . \\
  \end{array}
\end{equation}
The comparison between \eqref{eq:ELM-t+h} and
\eqref{eq:partitioned-method-t+h} shows that state $x$ in ELM is one
step ahead in terms of the differential variable.  Similar to
\eqref{eq:partitioned-method-chatter-free-t+ih},
\eqref{eq:ELM-chatter-free-t+ih} needs to hold for sufficient steps
starting from $k=1$ to exit chattering.
\begin{equation}
  \label{eq:ELM-chatter-free-t+ih}
  K_p \sum _{i=0}^k \Delta u_{t+ih} +h K_i (k u_t + \sum_{i=1}^{k} k\Delta u_{t+ih}) < 0 \, .
\end{equation}

\section{PI Controller Deadlock}
\label{sec:simultaneous}

\begin{figure}[!t]
\centering
\includegraphics[width=\columnwidth]{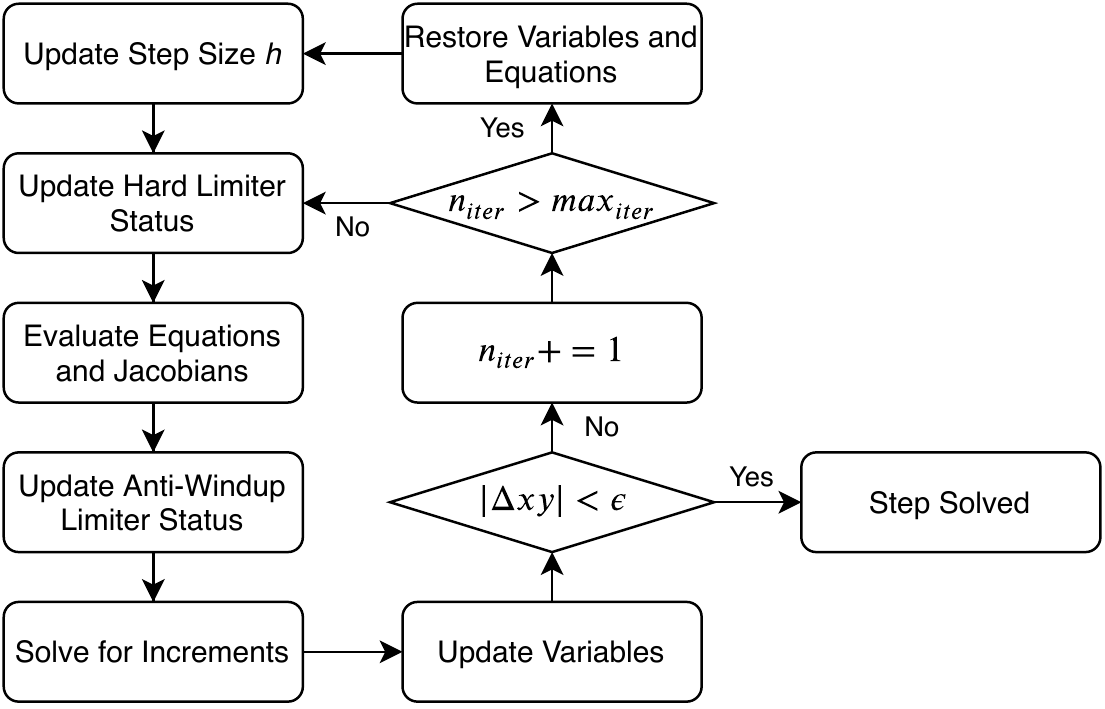}
\caption{The inner iterative integration loop for numerical integration.}
\label{fig:implicit-step}
\end{figure}

The PI controller may enter a deadlock state when simulated with ITM,
which implements an inner iteration loop. 
\figref{fig:implicit-step} shows the inner iteration loop for one integration step. 
Hard limiter status is based on the input and updated before equation evaluation.
Anti-windup limiter equations are dependent on both variables and equations and are thus updated after equation evaluation. 

Suppose at time $t-h$, the input and output satisfy conditions
\begin{equation}
\begin{array}{ll}
    \text{Input: } & u > 0, \dot{u} < 0 \\
    \text{Output: } &  \left\{\begin{matrix}
                                w = w_{max} \le y \\
                                z_u = 1, z_i = z_l = 0
                              \end{matrix}\right.
\end{array}{}
\label{eq:lock-condition-example}
\end{equation}

At time $t$, the integrator is disabled for the first iteration, and deadlock will happen if iterations meet the conditions:
\begin{itemize}
    \item If the input decreases to a value that renders $y < w_{max}$. Iterations will continue due to the increment for $y$. As a result, the AW will be enabled for the next iteration.
    \item Next, if the integrator output is so large that $y \ge w_{max}$. Iterations will continue due to the increment for $y$ and $w$. As a result, the AW will be disabled again.
\end{itemize}{}

In such a scenario, the iteration will continue until the maximum iteration number $N_{iter}$ is reached, without converging to a solution.
Deadlock of the controller occurs when the ITM fails to integrate for a given time step continuously. During the deadlock, the equations that get solved
switches between the two sets as given in \Cref{tab:conditional-equation-sets}.

\begin{figure}[!t]
  \centering
  \includegraphics[width=\columnwidth]{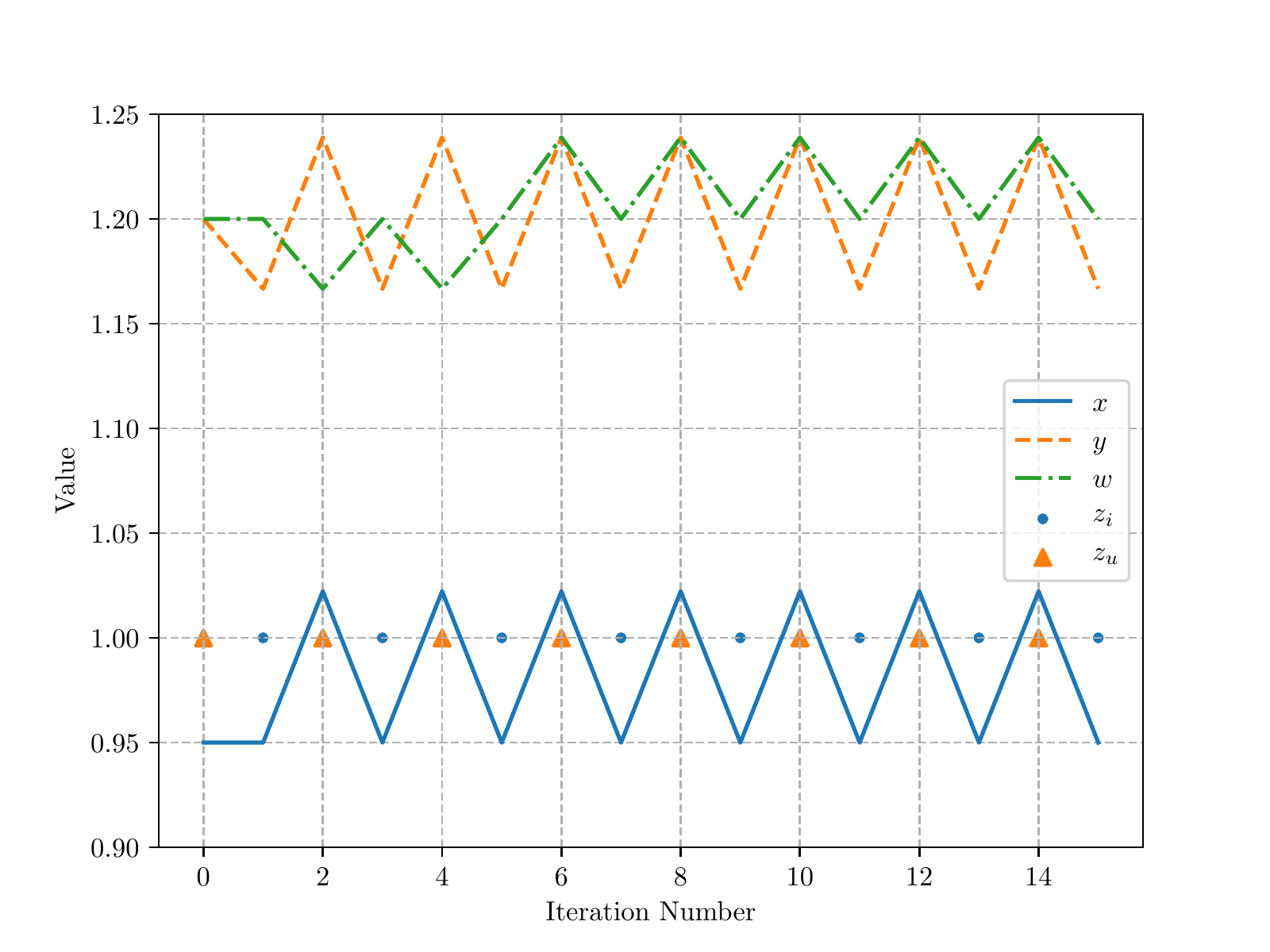}
  \caption{Illustration of a deadlock. Iteration number starts from
    zero, and values at the beginning of the corresponding iteration
    are plotted.}
  \label{fig:deadlock-illustration}
\end{figure}

\figref{fig:deadlock-illustration} illustrates a deadlock using a step
size of $10\, ms$.  The decrease in $u$ is reflected after iteration
zero, which unlocks the integrator for iteration one.  At iteration
one, the increment from the integrator is so large that $y$ exceeds
$w_{\max}$, causing the integrator to lock again for iteration two.
For each iteration, since the post-solution limiter status is
different from the pre-solution status, this process will continue
until $N_{\rm iter}$ is exhausted without
converging.

\subsection{Impact of Integration Step Size}

When non-convergence happens, an ITM solver can decrease the step
size based on predefined criteria \cite{milano2010power}. 
This variable step size approach is commonly used because software
can take advantage of the numerical stability of ITM by using a
larger size for most cases and shrink it when necessary.
The technique, however, does not
reduce the occurrence of deadlock.  Consider
a generic iteration $i$ of simulation time $t$, $y_t^{(i)}$ satisfies
\begin{equation}
  \begin{array}{ll}
    &y_t^{(i)} = K_p u_t + x_t^{(i)} \, , \\
    \text{where } &x_t^{(i)} = \left\{
    \begin{array}{ll}
    x_{t-h} &\text{if } z_{i, t} = 0 \, , \\
    x_{t-h} + \Delta x_{t}^{(i)} &\text{if } z_{i, t} = 1 \, ,
    \end{array}
    \right. 
\end{array}
\end{equation}
where $\Delta x_t^{(i)}$ is the increment for iteration $i$.
To avoid deadlock, if the integrator is unlocked at iteration $i$,
$y_t^{(i+1)}$ should stay below $w_{\max}$ to remain
the integrator unlocked. Therefore,
\begin{equation}
  \label{eq:simultaneous-iteration-i+1}
  \begin{array}{llll}
    y_t^{(i+1)} &= K_p u_t + x_t^{(i+1)} \\
          &= y_{t-h} + (K_p \Delta u_{t} + \Delta x_t^{(i+1)}) < w_{\max} \, , \\
      \Rightarrow & K_p \Delta u_t + \Delta x_t^{(i+1)} < 0 \, , 
  \end{array}
\end{equation}
applying ITM on \eqref{eq:simultaneous-iteration-i+1} and observing
that $\dot{x}_{t-h} = 0$, and $u_t > 0$, the step size needs to
satisfy \eqref{eq:step-size-to-avoid-deadlock-trapezoidal} to avoid
deadlock.
\begin{equation}
  \label{eq:step-size-to-avoid-deadlock-trapezoidal}
  h < - (2K_p / K_i) (\Delta u_t / u_t) \, .
\end{equation}

If we assume $u$ is the output of a low-pass filter and is thus
differentiable, applying ITM on \eqref{eq:simultaneous-iteration-i+1}
yields
\begin{equation}
  \label{eq:ITM-u-differentiable}
  h (\dot{u}_t + \dot{u}_{t-h}) < - 2(K_p/ K_i) (\dot{u}_t + \dot{u}_{t-h} ) - 2u_{t-h} \, .
\end{equation}
Note that $u$ decreases in the proximity of $t-h$, hence
$(\dot{u}_t + \dot{u}_{t-h}) < 0$. Dividing
\eqref{eq:ITM-u-differentiable} by $\dot{u}_t + \dot{u}_{t-h}$
yields
\begin{equation}
  \label{eq:ITM-h-condition-differentiable}
    h > -2 (K_p/ K_i) - 2 u_{t-h} / (\dot{u} + \dot{u}_{t-h}) \, .
\end{equation}

\eqref{eq:ITM-h-condition-differentiable} shows that the step size
needs to be greater than a minimum value to remain unlocked.
The variable step techniques, however, are designed to decrease
the step size when non-convergence happens.
This explains why in the case of deadlock, the simulation program
cannot improve the convergence by reducing the step size.
On the other hand, the integration step size has to be adequately
small for systems with fast dynamics.
As will be shown in \Cref{sec:simulation-verification}, the minimum
step size to avoid deadlock given in
\eqref{eq:ITM-h-condition-differentiable} can be too large to achieve.
Therefore, in some combinations of parameters and inputs, deadlock
could be inevitable because the step size condition cannot be
achieved.

\subsection{Impact of Convergence Tolerance}
Convergence tolerance also affects the deadlock in terms of the
iteration exit condition. As shown in \Cref{fig:implicit-step}, 
the inner Newton iteration will be deemed
converged if the maximum variable increment is smaller than the tolerance
$\epsilon$.
It explains why even if the step size in
\eqref{eq:step-size-to-avoid-deadlock-trapezoidal} is not
achievable, ITM can still converge.

Consider a deadlock scenario at time $t$ that after iteration $i-1$,
the integrator is unlocked.  The maximum increment, if we omit the
subscript $t$, is given by
\begin{equation}
  \max\left(  
    \left\{|\Delta x^{(i)}|, 
      |\Delta y^{(i)}|, 
      |\Delta w^{(i)}|
    \right\}
  \right) \, ,
\end{equation}
where, $|\Delta y^{(i)}| = |\Delta x^{(i)}|$, and
\begin{equation}
  \begin{array}{ll}
    \Delta w^{(i)} &= |y^{(i)} - w_{\max}| \\
    &\le |(w_{\max} + \Delta x^{(i)}) - w_{\max}| = |\Delta x^{(i)}| \, .
  \end{array}
\end{equation}
Applying ITM and observing $\dot{x}_{t-h} = 0$, $\Delta x_t^{(i)}$ is
given by
\begin{equation}
  \Delta x_{t}^{(i)} = 0.5 h \dot{x}_t^{(i)} \, ,
\end{equation}
where $\dot{x}_t^{(i)}$ ($i \in [0, N_{\rm iter}]$) toggles between $0$
and $K_i u_t$. The deadlock can exit only when the increment for
iteration $i$ is smaller than $\epsilon$, given by
\begin{equation}
  |\Delta x_{t}^{(i)}| = |0.5h K_i u_{t}| < \epsilon \, ,
  \label{eq:trapzoidal-increment}
\end{equation}
which indicates that, for a fixed step size, a temporarily lift in
tolerance can be implemented to stop the deadlock.  On the other hand,
given a fixed tolerance, \eqref{eq:trapzoidal-increment} allows
to calculate the maximum allowed (i.e., adequately small) step size to
stop deadlock with a variable step approach.

\section{Simulation Verification}
\label{sec:simulation-verification}

Numerical simulations in Simulink and \andes \cite{cui2020hybrid} are
performed to verify the analyses.  
Simulations are performed on a standalone PI controller as recommended
by the IEEE standard.
The PI controllers parameters are
from \cite{Murad2019} with $K_p=1$, $K_i=20$, and
\begin{equation}
  \dot{u} = 
  \begin{cases}
    -1 &  2< t < 6 \, , \\
    1 & \text{otherwise} .
  \end{cases}
\end{equation}
Substitute parameters into
\eqref{eq:partitioned-method-chatter-free-t+ih} and
\eqref{eq:ELM-chatter-free-t+ih} under step size $h=10^{-3}$ and
enumerate $k$ from 1 to 10, we can obtain the input value starting
from which the controller can remain unlocked.

\begin{figure}[t!]
  \centering
  \includegraphics[width=\columnwidth]{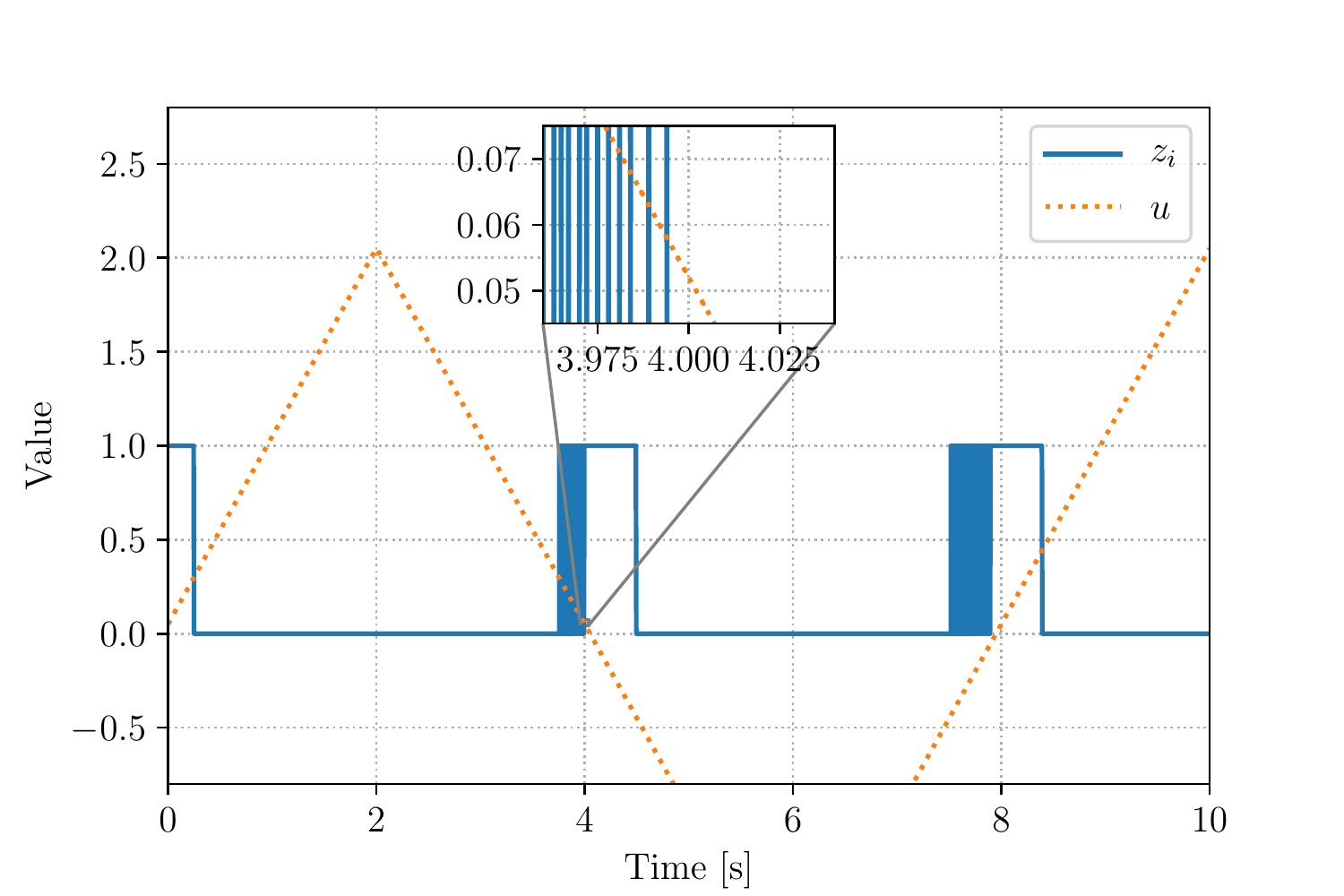}
  \caption{Simulink verification of the chattering stop condition.}
  \label{fig:simulink-ode1}
\end{figure}

Calculations show that for EPM and ELM, $u$ need to drop below
$0.0595$ and $0.0605$, respectively, to prevent the integrator from
switching back to the locked state.  Numerical integration in
Simulink™ using ODE1 verifies the calculation, as shown in
\Cref{fig:simulink-ode1}. It can be observed that the last relocking
happens when the input drops to $0.0605$. When the input further
decreases, the controller gets unlocked and stays unlocked thereafter.

For ITM, the initial step size $h$ and 
the convergence tolerance $\epsilon$ are both
set to $10^{-3}$.
A heuristic algorithm for adjusting the step size is employed.
The algorithm is based on the following rules:
\begin{equation}
  h_{t+h} =
  \begin{cases}
    h_t + 10^{-6} & \text{if } N_t \le 3 \, , \\ 
    h_t - 10^{-6} & \text{if } N_t \ge 15 \, , \\ 
    h_t & \text{otherwise} \, ,
  \end{cases}
  \label{eq:h-heuristic}
\end{equation}
where $N_t$ is the number of iterations taken to converge for
a generic time $t$.
The theoretical minimum step size to avoid deadlock from
\eqref{eq:ITM-h-condition-differentiable} is $h_{\min} = 0.1915$.
Compared with the cycle time of a 50/60 Hz power system, 
this step size is apparently too large for systems under disturbance.

\begin{figure}[!t]
\centering
\includegraphics[width=\columnwidth]{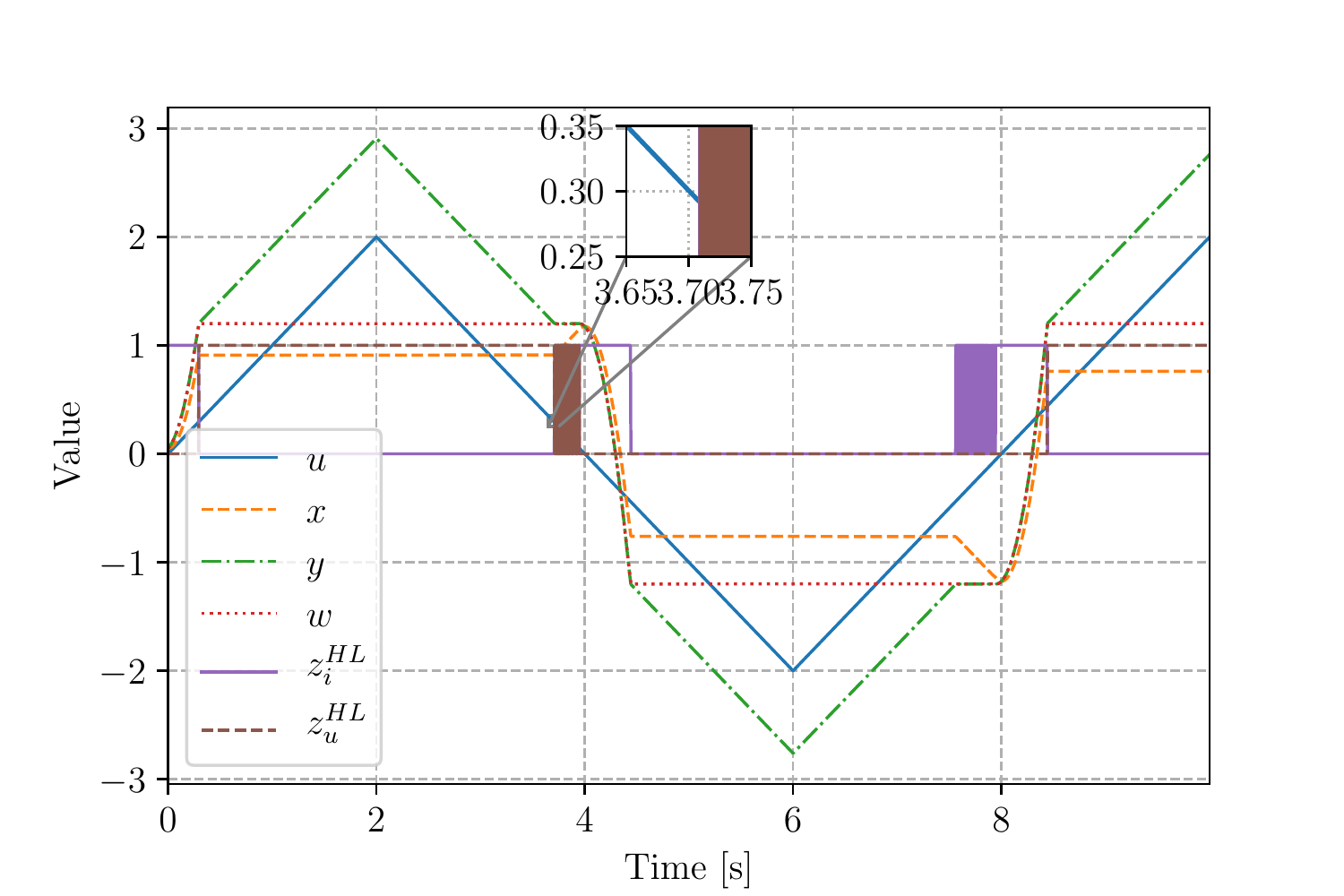}
\caption{PI controller deadlock and variable values using ITM.}
\label{fig:andes-values}
\end{figure}

Simulation results using \andes \cite{cui2020hybrid} 
are shown in \figref{fig:andes-values}. When the
deadlock begins at $t=3.709$ s, the input value $u_t=0.2915$.

\begin{figure}[t!]
  \centering
  \includegraphics[width=\columnwidth]{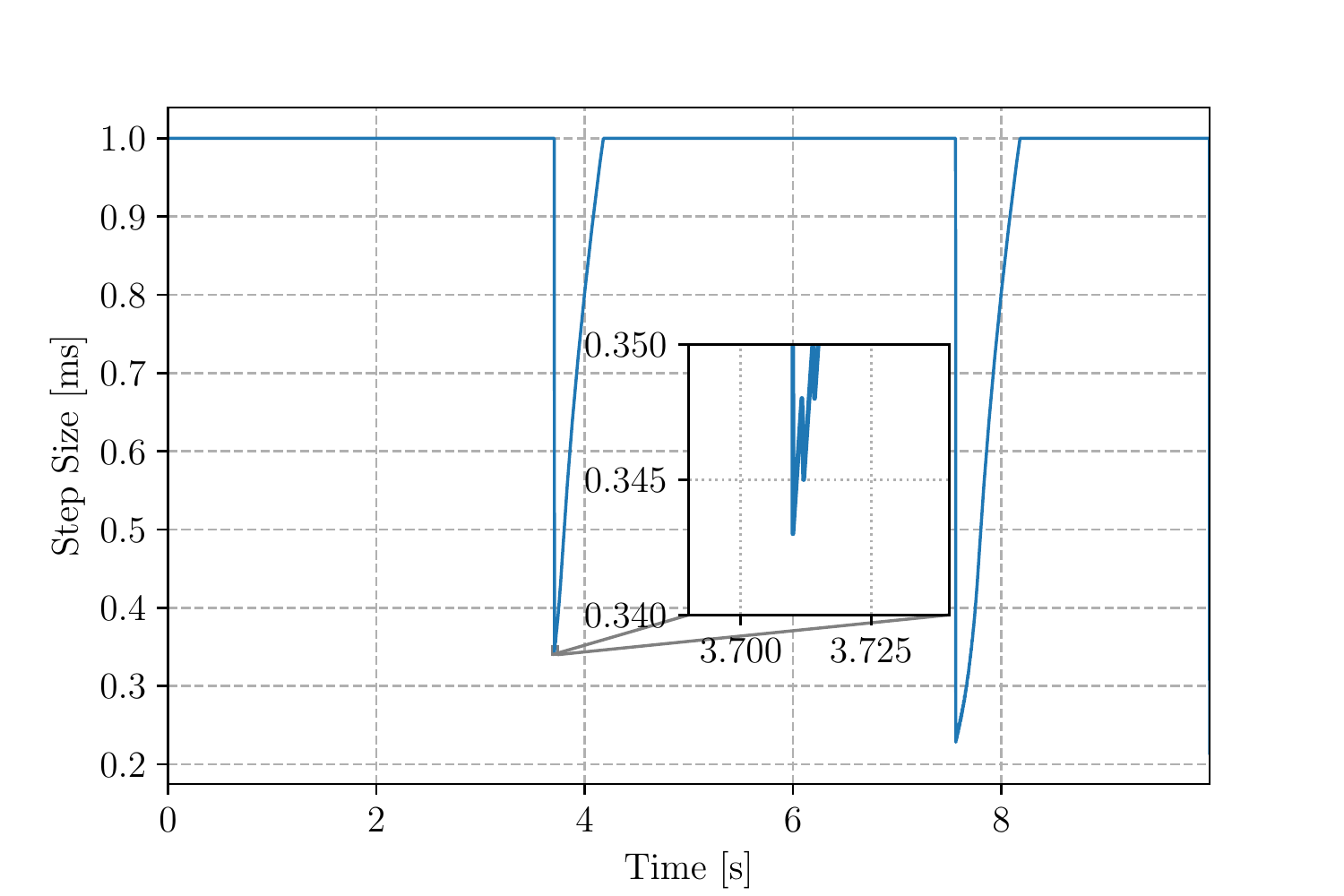}
  \caption{Integration step size adjustments using ITM.}
  \label{fig:andes-h}
\end{figure}

Next, Using \eqref{eq:trapzoidal-increment} and a step size cap of $10^{-3}$, the step
size needs to satisfy $h < 0.3431$ ms to exit deadlock.
\Cref{fig:andes-h} shows the step size change and verifies the
calculation.

\balance

\section{Conclusions}
\label{sec:Conclusions}
This paper investigates the chattering and deadlock issue of the PI
controller recommended by the IEEE Standard 421.5-2016 under three most
commonly used numerical integration methods.
For the non-iterative EPM and ELM, the chattering issue is discussed with
the chattering stop conditions deduced, respectively.
For the iterative ITM, the deadlock caused by non-convergence of the inner
Newton iteration loop is explained.
The impacts of step size and convergence tolerance
on the ITM deadlock are also discussed.

The most interesting conclusion from the analysis is that, for some
combinations of parameters and inputs, deadlock is inevitable
since the step size requirement to avoid deadlock cannot be achieved. 
However, after shrinking the step size using the variable step approach,
ITM will exit the deadlock once the convergence tolerance is satisfied.

\ifCLASSOPTIONcaptionsoff
  \newpage
\fi

\bibliographystyle{IEEEtran}
\bibliography{IEEEabrv,papers}

\end{document}